\documentclass[aps,pra,superscriptaddress,twocolumn,10pt]{revtex4-2}

\usepackage[utf8]{inputenc}
\usepackage[version=3]{mhchem}
\usepackage{amsmath}
\usepackage{amsfonts}
\usepackage{hyperref}
\usepackage{braket}
\usepackage{lipsum}
\usepackage{url}
\usepackage[draft]{graphicx}
\usepackage{lipsum}
\usepackage{color}
\usepackage{float}
\usepackage[export]{adjustbox}
\usepackage{wrapfig}
\usepackage{array,multirow}
\usepackage{tabularx}
\usepackage{bm}
\usepackage{xfrac}

\makeatletter
\renewcommand\frontmatter@abstractwidth{\dimexpr\textwidth-1in\relax}
\makeatother

\begin{abstract}
We introduce high-harmonic sideband spectroscopy (HHSS) and show that it can be a robust probe of charge migration (CM) in a halogenated carbon-chain molecule. We simulate both the CM and harmonic-generation (HHG) dynamics using {\it{ab initio}} time-dependent density-functional theory. We find that CM dynamics initiated along the molecular backbone induces sidebands in the HHG spectrum driven by a delayed laser pulse that is polarized perpendicular to the molecular axis. Monitoring the spectrum as either the HHG laser frequency or the relative delay is scanned allows for the extraction of detailed information about the time-domain characteristics of the CM process. 
\end{abstract}

\begin{document}

\author{Kyle~A.~Hamer}
\email{khamer3@lsu.edu}
\affiliation{Department of Physics and Astronomy, Louisiana State University, Baton Rouge, LA 70803, USA}

\author{Fran\c{c}ois~Mauger}
\affiliation{Department of Physics and Astronomy, Louisiana State University, Baton Rouge, LA 70803, USA}

\author{Aderonke~S.~Folorunso}
\affiliation{Department of Chemistry, Louisiana State University, Baton Rouge, LA 70803, USA}

\author{Kenneth~Lopata}
\affiliation{Department of Chemistry, Louisiana State University, Baton Rouge, LA 70803, USA}
\affiliation{Center for Computation and Technology, Louisiana State University, Baton Rouge, LA 70803, USA}

\author{Robert~R.~Jones}
\affiliation{Department of Physics, University of Virginia, Charlottesville, VA 22904, USA}

\author{Louis~F.~DiMauro}
\affiliation{Department of Physics, The Ohio State University, Columbus, OH 43210, USA}

\author{Kenneth~J.~Schafer}
\affiliation{Department of Physics and Astronomy, Louisiana State University, Baton Rouge, LA 70803, USA}

\author{Mette~B.~Gaarde}
\email{mgaarde1@lsu.edu}
\affiliation{Department of Physics and Astronomy, Louisiana State University, Baton Rouge, LA 70803, USA}

\title{Characterizing Particle-Like Charge Migration Dynamics\\ with High-Harmonic Sideband Spectroscopy
}
\date{\today}
\maketitle


\section{Introduction}

Electron dynamics at the attosecond timescale are the first stage in the molecular response to ultrafast photoexcitation or ionization \cite{Breidbach2005, Remacle2006, Lepine2014}. One example, which has attracted intense interest in the ultrafast community, is charge migration (CM): this coherent, back-and-forth hole motion is often considered a precursor to more permanent structural or chemical changes \cite{Cederbaum1999, Lunnemann2008, Calegari2014, Kraus2015, Golubev2015, Calegari2016, Kuleff2016, Ayuso2017, Bruner2017, Worner2017, Lara-Astiaso2018, Mansson2021}. A number of theoretical and experimental studies concerning CM have been performed: for example, documenting and characterizing CM \cite{Lunnemann2008, Leeuwenburgh2013, Calegari2014, Kraus2015, Golubev2015, Calegari2016, Kuleff2016, Ayuso2017, Bruner2017}, investigating how to control CM and how it depends on the specifics of the molecule \cite{Golubev2015, Despre2019, folorunso2021}, and more recently, the evolution of the CM as the electron dynamics couple to the nuclear degrees of freedom \cite{Vacher2017, Lara-Astiaso2018, Despre2018, Mansson2021}.

Due to the challenges of experimental control at the attosecond timescale, CM has only been detected in a handful of measurements \cite{Calegari2014, Kraus2015, Calegari2016, Lara-Astiaso2018, Mansson2021}. The majority of these have documented a modulation in the double ionization signal of a molecular fragment following ultrafast photoionization. These delay-dependent variations have been ascribed to CM, with theory showing similar modulations in the populations of cationic states \cite{Calegari2014, Calegari2016, Ayuso2017, Lara-Astiaso2018, Mansson2021}. Only a single experimental study employing self-probing high-harmonic spectroscopy (HHS) has been published. In that experiment the ionization and recombination steps of the high-harmonic generation (HHG) process were used to initiate and probe the CM, respectively \cite{Kraus2015}. Accordingly, the CM dynamics were re-initiated in each half-laser cycle and did not manifest as a straightforward modulation \cite{Calegari2014}, but was interpreted with the help of model calculations \cite{Kraus2015}. A number of experimental studies are currently ongoing \cite{Matselyukh2020, Barillot_PRX2021,Li2022}, many employing pump-probe techniques and exploring different observables through which the CM might be characterized. 

\begin{figure}[tb]
    \centering
    \includegraphics[width=\linewidth]{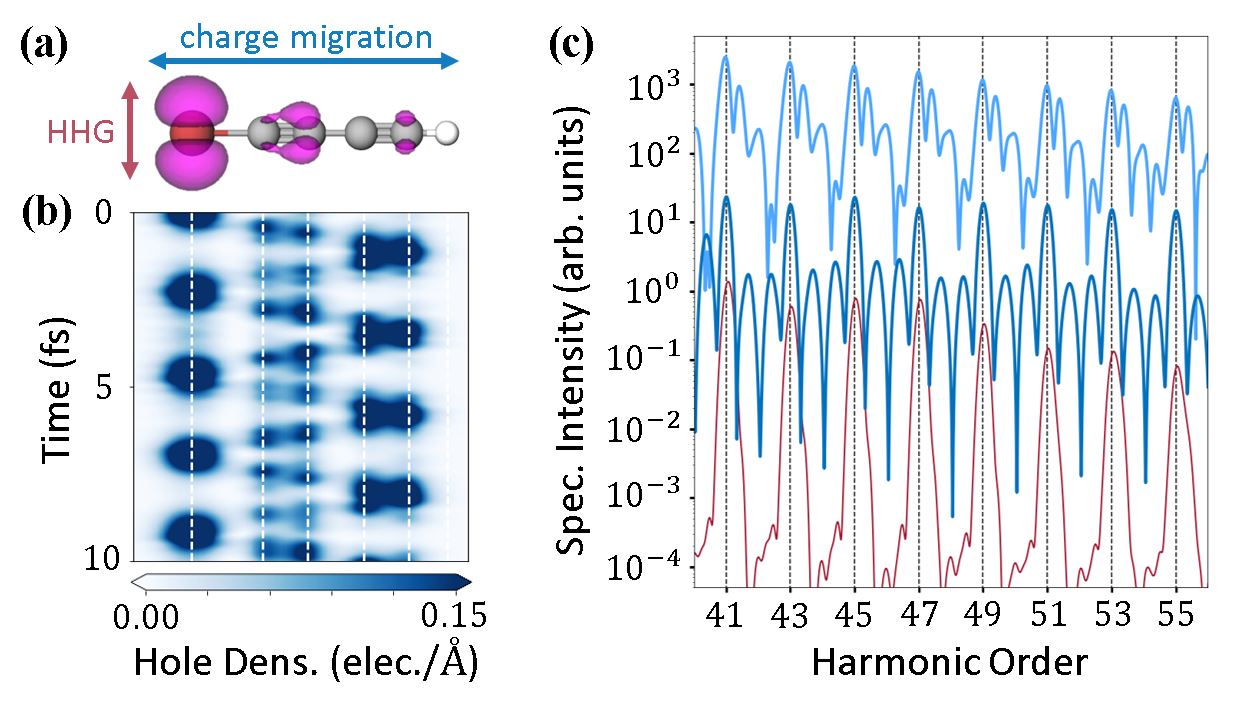}
    \caption{(a) Schematic of HHSS configuration to probe CM in BrC$_{4}$H. The HHG driving field is polarized perpendicular to the molecular backbone along which the periodic CM occurs. We also show the isosurface of the electron density contribution from the unpaired Kohn-Sham channel in which we introduce the initial localized-hole perturbation.
    (b) Time evolution of the electron density contribution from (a), integrated over the directions transverse to the molecular backbone, as a function of position.
    (c) Multi-cycle HHG spectra for 1575-nm (top light blue) and 1800-nm (middle dark blue) driving wavelengths. The bottom thin red curve shows the 1800-nm spectrum in the absence of CM (in the neutral molecule). For clarity, the light blue curve is offset by a factor of 200.} 
    \label{fig:1}
\end{figure}

In this paper, we introduce high-harmonic sideband spectroscopy (HHSS), a robust probe of CM that uses HHS as an independent probe step following a CM-initiating pump. As illustrated in Fig.~\ref{fig:1}(a), the laser field driving the HHG is polarized perpendicular to the molecular backbone, to avoid driving the CM dynamics \cite{Kraus2015}. We use TDDFT to describe both the CM and the HHG dynamics, and directly calculate the high-harmonic spectrum observable in experiment. We show that periodic CM dynamics along the backbone of a molecule, initiated by the creation of a localized hole, leads to coherent modulation of the time-dependent HHG yield produced by a delayed probe pulse, and we demonstrate how to extract detailed information about the time-domain characteristics of the CM motion. In the time domain, the modulation arises as the beat between the incommensurate frequencies of the CM motion and the HHG driving field, and reflects the sensitivity of the HHG process to the space- and time-dependence of the molecular charge density. In the frequency domain, the modulation manifests as sidebands whose energies correspond to the sum and difference frequencies of the probe-generated harmonics and the CM. These sidebands are a background-free probe of the CM motion and we show how one can unambiguously extract the CM frequencies by scanning the driving laser wavelength. More broadly, we discuss how the nature of the CM dynamics is imprinted on the HHSS signal, both in the spectral and temporal domains. Finally, we discuss the viability of HHSS for experimental implementation, e.g. the inclusion of a distribution of molecular alignment angles and the influence of decoherence.

The paper is organized as follows. Section II describes our theoretical and numerical approach, detailing the process of using TDDFT to simulate both the CM and HHG. Section III presents results of wavelength and delay scans which explore the modulation of the HHG process via the CM, and explores the experimental viability of our approach. Lastly, in Section IV, we summarize and discuss the prospects of our results. Unless otherwise stated, atomic units (a.u.)\ are used throughout the paper.


\section{Methodology}

Figure~\ref{fig:1}(a) illustrates our general approach for performing HHSS of CM, using bromobutadiyne (BrC$_{4}$H) as an example. We start by creating a one-electron valence hole localized on the Br center, which induces the periodic particle-like CM dynamics shown in Fig.~\ref{fig:1}(b) with a frequency of $\omega_\text{CM}\approx1.84$~eV. In the calculations, this localized hole is achieved by using constrained density functional theory~\cite{eshuis2009}, as outlined in~Appendix~\ref{sec:init_cond} and in Ref.~\cite{folorunso2021}. This localized initial hole emulates a sudden core or inner-valence ionization, localized on Br, that could be created by an attosecond x-ray pump pulse~\cite{Barillot_PRX2021}, subsequently leading to CM in the valence shell of the molecule~\cite{Cederbaum1999, Bruner2017, folorunso2021, Mauger2022}. Next, a delayed mid-infrared (MIR) laser pulse, polarized perpendicular to the molecular backbone, drives the HHG-probe in the molecular cation as it undergoes CM.

In order to describe both the HHG and CM processes, we use grid-based time-dependent density-functional theory (TDDFT) \cite{kohn1965} with a local-density-approximation exchange-correlation functional \cite{Bloch1929, dirac1930, perdew1981, marques2012} and an average-density self-interaction correction \cite{legrand2002} within Octopus \cite{andrade2012, andrade2015, castro2004}. Nuclear dynamics are omitted in all TDDFT results shown in this paper; however, in Section \ref{sec:3C} we discuss the effect of decoherence, {\it{e.g.}}, due to coupled electron-nuclear motion, on the effectiveness of HHSS. Technical details about how the TDDFT simulations are performed can be found in Appendix~\ref{sec:TDDFT_sims}.

In all simulations, we ramp up the MIR field amplitude over two full laser cycles, followed by eight cycles with constant amplitude at an intensity of $60~\text{TW}/\text{cm}^{2}$. To select the short-trajectory contribution that is usually observed in HHG measurements~\cite{bellini1998, antoine1996}, we use a combination of a weak attosecond-pulse-train (APT) ionization-seed synchronized with the MIR~\cite{schafer2004} and a simulation box which is scaled to the quiver radius in the laser polarization direction \cite{hamer2021}. Finally, we compute harmonic spectra from the total dipole acceleration in the direction of the laser polarization. We discard the ramp-up part of the signal to avoid the transient effects associated with the initial interaction between the driving laser and the CM dynamics. We have checked that including the ramp-up contribution and/or using the total harmonic spectrum (sum of the parallel and perpendicular components) has only cosmetic effects and does not change our results or conclusions. We show examples of HHG spectra, both with and without CM, in Fig.~\ref{fig:1}(c). The CM-active molecule is highly excited and, in simulations, we find that its HHG yield is overall comparable to that of the neutral ground state (even in the absence of the APT).

Without pump-initiated CM -- thin red curve in Fig.~\ref{fig:1}(c) -- we observe well-resolved odd-harmonic peaks, as expected. By comparison, the solid blue curves showing the HHG spectra with CM for two different laser wavelengths exhibit additional peaks located between the harmonic orders. These {\it sidebands} are caused by the modulation of the HHG process due to the CM motion along the molecular backbone. This periodic modulation gives rise to a beat in the time-dependent harmonic yield since the laser and CM frequencies $\omega_L$ and $\omega_{\text{CM}}$ are, in general, incommensurate. As illustrated in the figure, the location, strength, and number of sidebands depend on the driving laser frequency relative to the CM frequency. A central result of this paper is that these sidebands, and more generally, the underlying CM-driven modulation of the HHG, are a clear and background-free probe of the CM dynamics. Note that any potential MIR-triggered CM is re-initiated every half laser-cycle and therefore contributes only to the signal at the odd-harmonic frequencies. In what follows, we discuss how we extract detailed information about the period and the nature of the CM dynamics from the modulated harmonic spectra. Also, we note that the sidebands that are present in Fig.~\ref{fig:1}(c),  and in Figs. 2 and 3 below, are still visible when accounting for imperfect molecular alignment. We will return to this in more detail in Section~\ref{sec:3C}. 


\section{Results}

\subsection{Wavelength Scan}\label{sec:3A}

\begin{figure}[tb]
    \centering
    \includegraphics[width=0.9\linewidth]{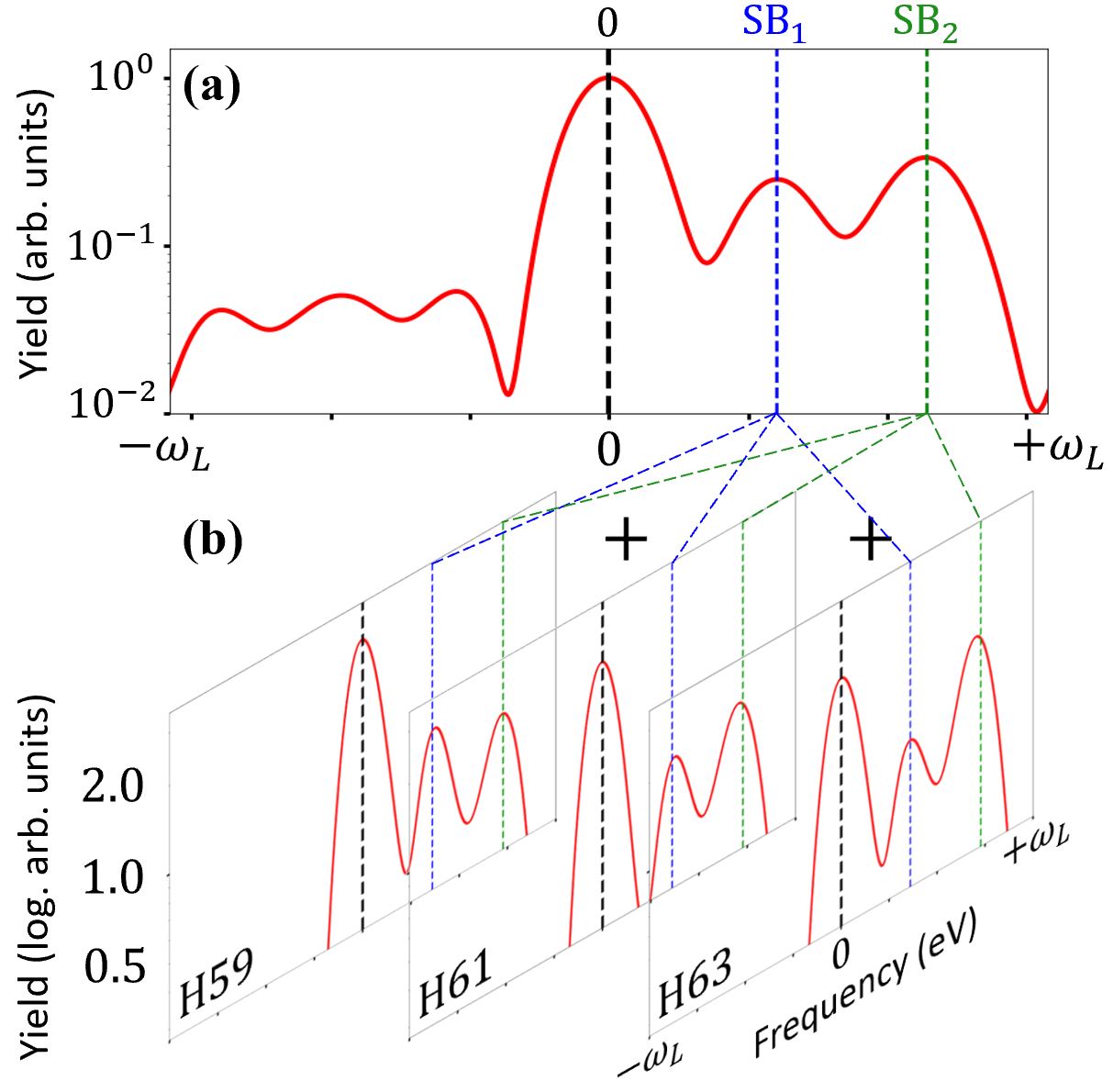}
    \caption{(a) Stacked HHG spectrum around one odd harmonic at 1575 nm. Sideband peaks 1 and 2 are marked. (b) Schematic of the stacking process we use to produce (a), where we average $2\omega_{L}$-wide slices of the relevant HHG spectrum centered around successive odd harmonics. Here, we have adjusted the $y$-axis such that the noise floor is not shown.}
    \label{fig:2}
\end{figure}

We start our sideband analysis in the frequency domain, by showing that the energies of these sidebands can be used to directly measure the CM period. To that end, we track the energy separation between the sideband and odd harmonics, as illustrated in Fig.~{\ref{fig:2}}(a) for a driving wavelength of 1575~nm. Since the sideband amplitudes vary considerably across the harmonic spectrum, we use a stacked spectrum obtained by averaging over $2\omega_{L}$-wide slices of the entire spectrum above 20~eV. The stacking process is illustrated in Fig.~\ref{fig:2}(b) for the strips associated with harmonics~57 to 61. The stacked spectrum in Fig.~\ref{fig:2}(a) clearly shows two sideband peaks to the right of the central odd harmonic, marked by $\text{SB}_{1}$ and $\text{SB}_{2}$. The frequencies of $\text{SB}_{1}$ and $\text{SB}_{2}$ result from the sum of a harmonic frequency and $\omega_{\text{CM}}$ or 2$\omega_{\text{CM}}$, so that the characteristic $\text{SB}_{1}$ and $\text{SB}_{2}$ frequency differences in the spectrum are given by
\begin{equation}\label{eq:freqdiffs}
    \Delta \omega_{1} = \pm (\omega_{\text{CM}} - 2\omega_{L}) \quad \text{and} \quad 
    \Delta \omega_{2} = 2 \Delta \omega_{1}.
\end{equation}
It is worth noting that there is nothing special about the sidebands on the high-energy side of the harmonic peaks, and that there are other driving wavelengths for which the low-energy sidebands dominate. This fluctuation in the relative strength between the harmonics and the sidebands is due to our finite sampling of the beat between incommensurate frequencies. 

The fact that {\it two} sidebands are present above (and below) each harmonic is significant: it means that the harmonics are modulated not just at $\omega_{\text{CM}}$ but also at 2$\omega_{\text{CM}}$. This modulation is consistent with the particle-like CM shown in Fig.~{\ref{fig:1}}(b), in which the time-dependent hole density evolves at both the fundamental CM frequency and its second harmonic. In this time-domain picture, the 2$\omega_{\text{CM}}$-modulation of the density dynamics is most clearly visible around the central pair of carbon atoms, where the hole passes through twice per CM period (we will return to how this manifests in the frequency domain in Section \ref{sec:3B}). The presence of both sets of sidebands thus means the HHSS probe is sensitive to the full dynamical evolution of the CM. 

\begin{figure}[tb]
    \centering
    \includegraphics[width=\linewidth]{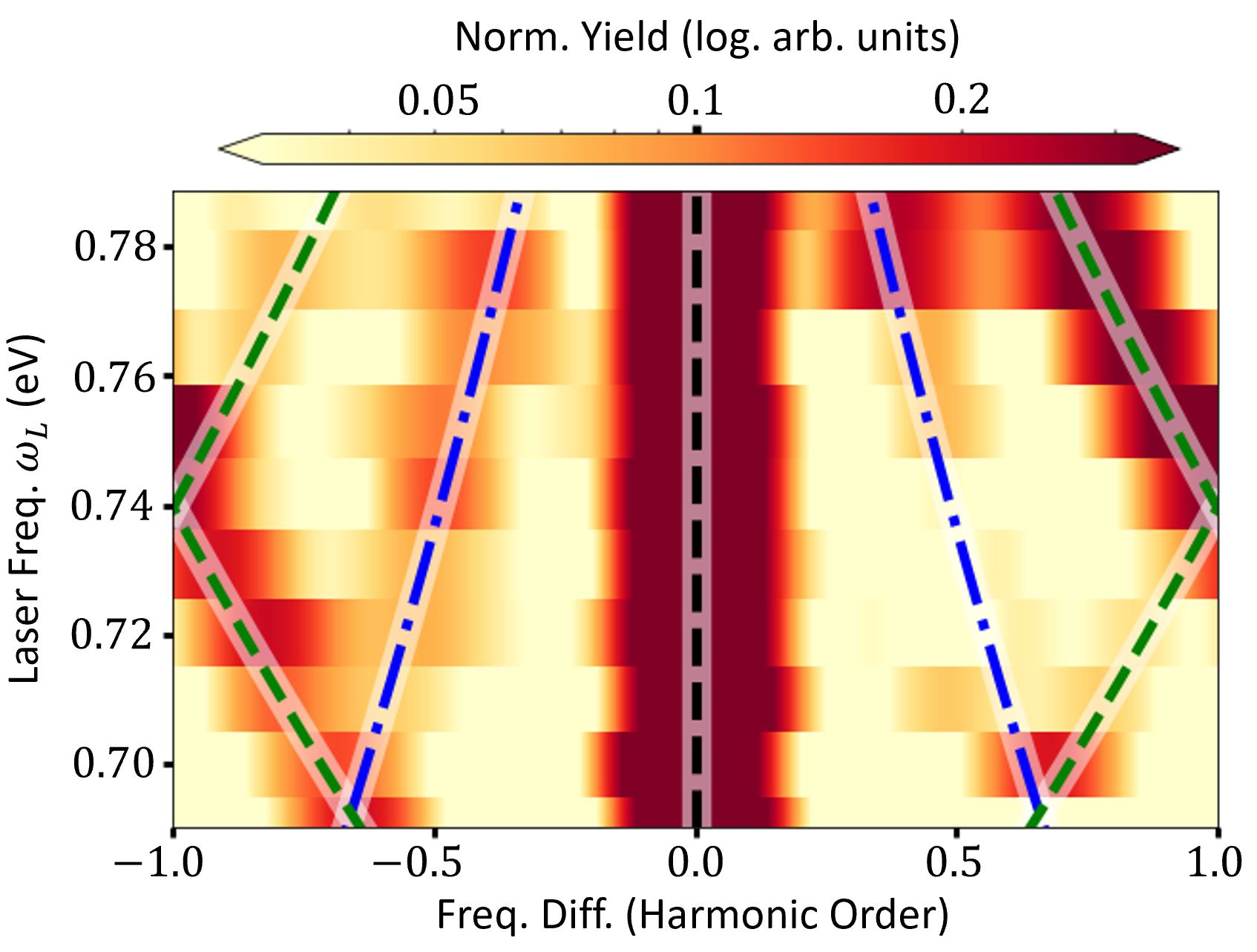}
    \caption{Stacked HHG spectra as a function of laser frequency, for laser wavelengths between 1575 nm and 1800 nm. The trendlines correspond to the sum and difference frequencies $\Delta \omega_{1}$ and $\Delta \omega_{2}$ of Eq.~(\ref{eq:freqdiffs}), as fitted in Fig~\ref{fig:4}(b).}
    \label{fig:3}
\end{figure}

Next, we show that the accuracy of the extraction of $\omega_{\text{CM}}$ increases if one scans the driving laser frequency. This is illustrated in the contour plot in Fig.~{\ref{fig:3}} that shows clear patterns in the evolution of the sideband location relative to the central odd harmonic. The four trendlines correspond to the sum and difference frequencies $\Delta \omega_{1}$ and $\Delta \omega_{2}$ of Eq.~(\ref{eq:freqdiffs}), highlighted by the blue and green lines, respectively. The green dashed lines have a discontinuity near $\omega_{L} = 0.74~\text{eV}$ (1675~nm), when $\text{SB}_{2}$ reaches the edge of the stacking box and reemerges on the other side of it. This means that sidebands from different harmonics interfere with each other at the edges of the stacked spectra, and then cross. Similarly, for laser frequencies near $\omega_{L} = 0.70~\text{eV}$ ($1775~\text{nm}$), $\text{SB}_{1}$ and $\text{SB}_{2}$ cross paths and interfere with each other. Finally, note that we are mostly interested in the location of the stacked-spectrum sidebands here, since the relative strengths of sidebands below or above the harmonic peak changes with wavelength, due to our finite sampling of the incommensurate-frequencies.
\begin{figure}[tb]
    \centering
    \includegraphics[width=\linewidth]{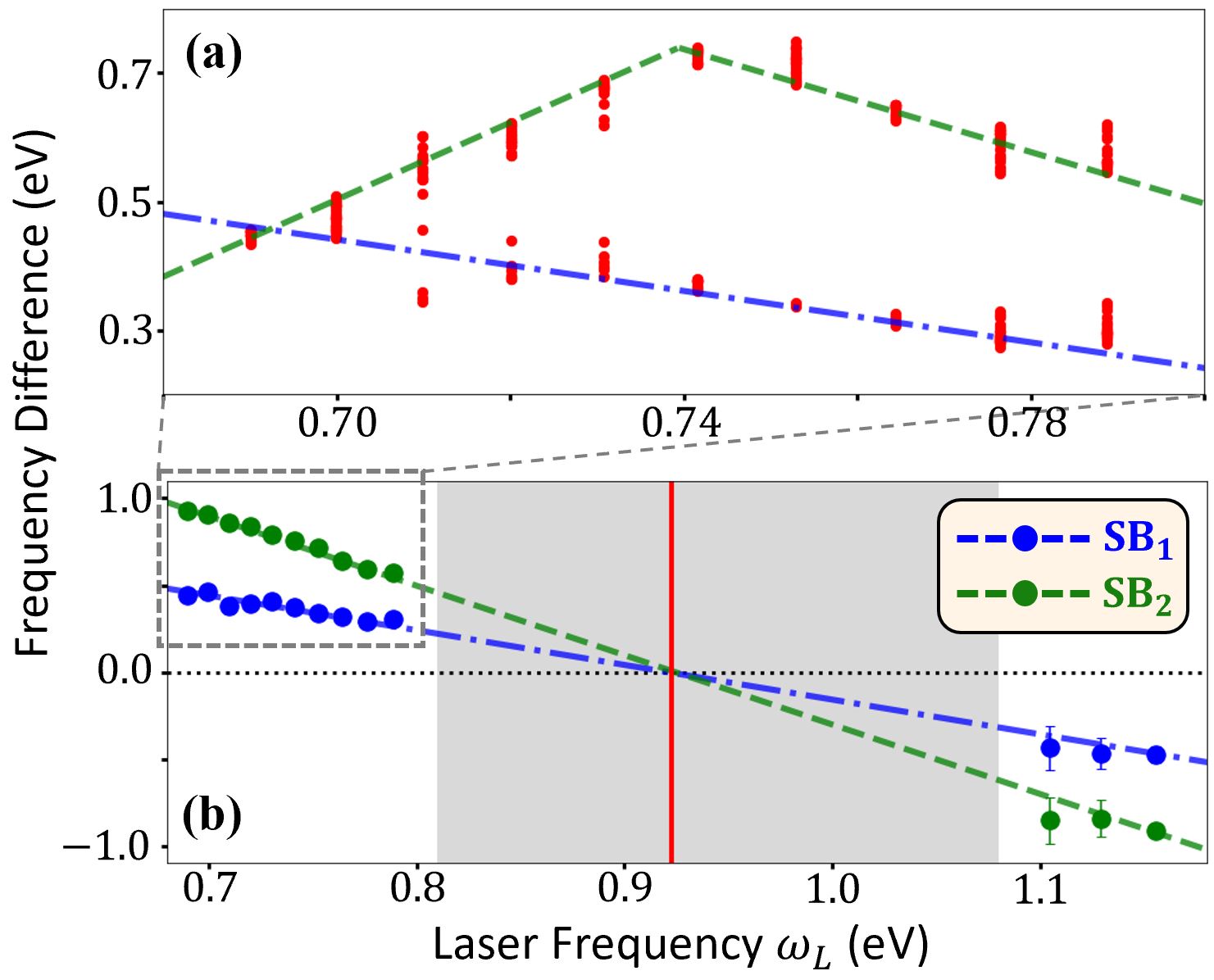}
    \caption{(a) Individual harmonic and sideband peaks analysis. Markers show the frequency difference between all the sideband peaks and their nearest odd harmonic throughout the HHG plateau for driving laser wavelengths between 1575 and 1800~nm. The two dashed curves mark the linear regression of the clusters of points associated with $\text{SB}_{1}$ and $\text{SB}_{2}$ (see text). (b) Sideband analysis to extract the CM frequency. The markers show the average energy separation of SB$_1$ and SB$_2$ from their odd harmonics with the error bars marking the standard deviation -- note that most error bars are contained within their respective markers. The vertical red line marks $\omega_{CM}/2$, where SB$_1$, SB$_2$, and the odd harmonics are all predicted to overlap.}
    \label{fig:4}
\end{figure}

To extract $\omega_{\text{CM}}$ from the frequency scan, we systematically record the frequency difference between the sidebands and their nearest odd harmonics across the plateau and sort these peaks into two groups associated with $\text{SB}_{1}$ and $\text{SB}_{2}$. To do so, we record the energies of all of the peaks in the HHG intensity within the harmonic plateau, and then sort these peaks into harmonics (at odd multiples of the MIR frequency) and sidebands (other peaks). In Fig.~\ref{fig:4}(a), we plot the frequency difference between each of the thus-detected sideband peaks and their closest odd harmonic for driving wavelengths between 1575 and 1800~nm. Clearly, in the figure, all the sideband peaks fall into two groups associated with $\text{SB}_{1}$ and $\text{SB}_{2}$. For each wavelength, we automate the sorting of the sideband peaks between $\text{SB}_{1}$ and $\text{SB}_{2}$ using a  $k$-means clustering algorithm~\cite{wang2011}, with $k=2$ since there are two sets of sidebands. Compared to the stacked spectrum, the analysis of the individual peaks across the entire harmonic plateau provides not only the average sideband energies $\Delta \omega_{1}$ and $\Delta \omega_{2}$ (which are consistent with the stacked spectrum results) but also a measure of the corresponding uncertainties.

In Fig.~{\ref{fig:4}}(b), we plot the mean value and standard deviation of the sideband energy difference as functions of the laser frequency $\omega_{L}$. For clarity, we have unfolded SB$_2$ below $\omega_L=0.74$~eV to avoid the discontinuity seen in panel~(a). Note that this unfolding does not require prior knowledge of the CM frequency, only that of the driving laser. We have also chosen to plot the frequency differences with positive (negative) values for laser wavelengths where $2\omega_{L} < \omega_{\text{CM}}$ ($2\omega_{L} > \omega_{\text{CM}}$). Again, this does not require a prior knowledge of the CM frequency since it can be inferred from the sign of the slopes in the stacked spectrum like in panel~(b). The shaded gap in the middle marks the frequencies where the sideband peaks become indistinguishable, within our resolution, from the harmonics in the HHG spectrum. To recover the CM frequency, we then fit the sideband energies against the prediction of Eq.~(\ref{eq:freqdiffs}). For both data sets, the fits are excellent. Each set of sidebands provides an independent measurement of $\omega_{\text{CM}}/2$ with
\begin{equation} \label{eq:measuredCMfreq}
\begin{array}{l}
    \text{SB}_{1} :\ \omega_{\text{CM}} = 1.841 \pm 0.004\ \text{eV},\\
    \text{SB}_{2} :\ \omega_{\text{CM}} = 1.849 \pm 0.002\ \text{eV}.
\end{array}
\end{equation}
The extracted frequencies of Eq.~(\ref{eq:measuredCMfreq}) are both in good agreement with each other and also consistent with the value $1.83 \pm 0.05$~eV we obtain directly from the CM dynamics in Fig.~\ref{fig:1}(b). We note that the inclusion of laser frequencies above {\it and} below the CM frequency greatly improves the precision of the measurement of the CM frequency $\omega_{\text{CM}}$.


\subsection{Temporal Analysis}\label{sec:3B}

\begin{figure}[tb]
    \centering
    \includegraphics[width=\linewidth]{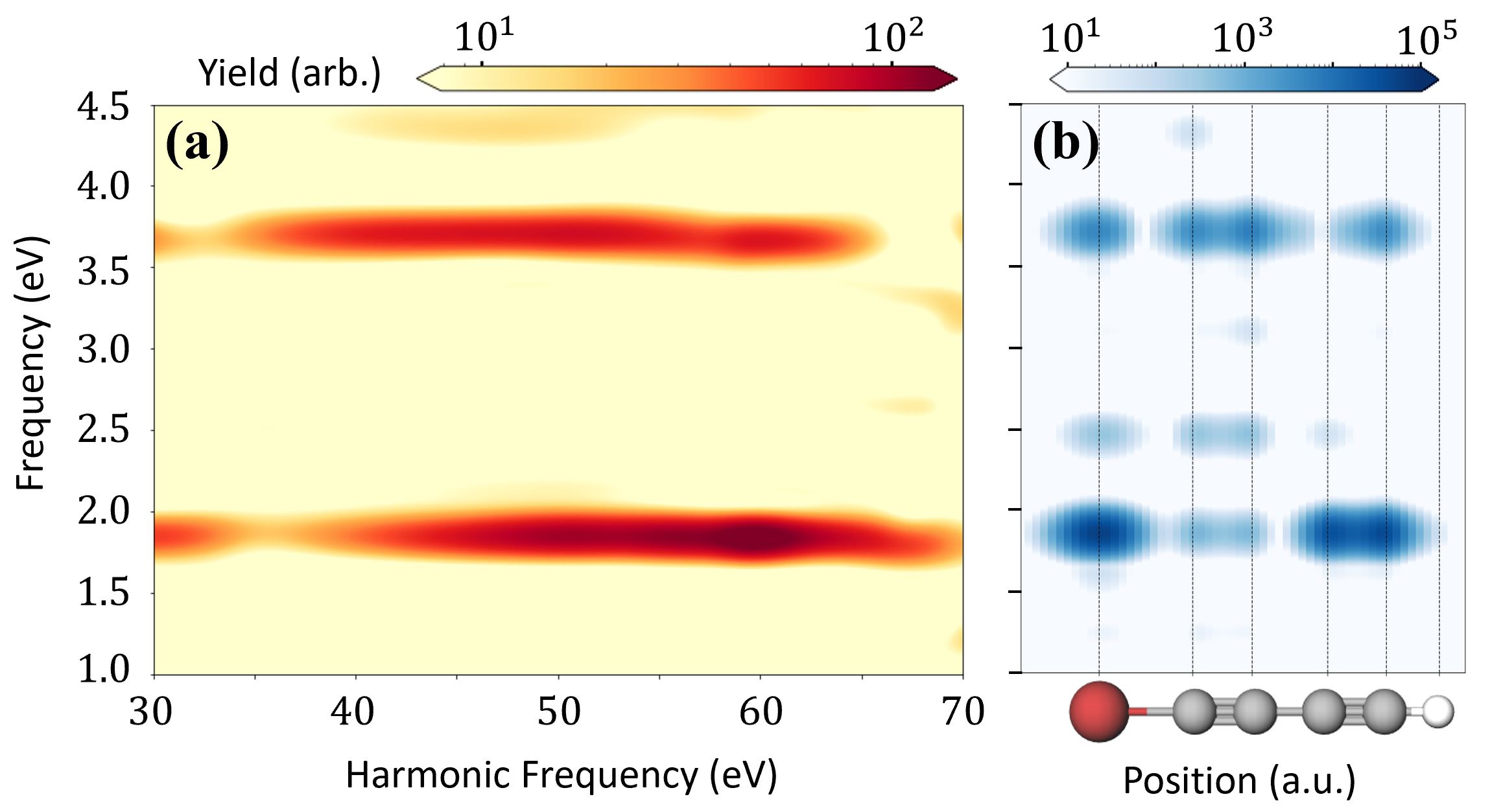}
    \caption{Frequency-domain representation of the time-dependent CM dynamics: (a) Fourier transform of the collated time-dependent harmonic spectra at 1575~nm (see text), as a function of harmonic energy. 
    (b) Fourier transform calculated directly from the time-dependent hole density of Fig.~{\ref{fig:1}}(b), as a function of the position along the molecular backbone.}
    \label{fig:5}
\end{figure}

Additional characteristics of the CM dynamics are imprinted on the HHSS signal, in the form of the relative strength of the $\omega_{\text{CM}}$ and $2\omega_{\text{CM}}$ contributions. As seen in Fig.~\ref{fig:3}, there are many laser wavelengths for which the two sidebands interfere either with each other or with an odd harmonic, even with the multi-laser-cycle duration of the CM signal we are considering here. This hinders a quantitative analysis of the relative strength of the sidebands. As an alternative, we consider the CM-induced modulation of the HHSS signal directly in the time domain, similar in spirit to \cite{Kraus2013}. 
Fig.~\ref{fig:5}(a) provides a time-resolved analysis of the modulation of the HHG signal by the CM dynamics. For this we combine the decomposition of the half-cycle-resolved HHG signal \cite{hamer2021} for various CM-MIR delays. Specifically, in the multi-cycle dipole acceleration signal we isolate the individual HHG contribution from each half-laser-cycle using a $\sfrac{1}{2}$-cycle-duration $\cos^{4}$ window centered around the corresponding extremum of the laser field. For each TDDFT simulation, this provides a coarse time-resolved HHG signal with a sampling resolution of one-half laser cycle. We then refine this sampling resolution by collating the results of 8 half-cycle-resolved HHG spectra for various delays between the CM and MIR (equally spaced within $0\leq\Delta<0.5$~o.c.), such that the collated time-resolved HHG spectrum has a sampling resolution of $\sfrac{1}{16}$ optical cycles. Then we obtain Fig.~\ref{fig:5}(a) by Fourier transforming this collated HHG signal, using a $\cos^{2}$ window that spans the entire time signal, with respect to this sub-cycle delay. In an experiment, this would correspond to using a short CEP-stabilized generation pulse with a controllable delay relative to the initiation of the CM, and scanning the delay with sub-cycle resolution over several laser cycles (unlike an experiment to look at sidebands, for which sub-cycle control of the delay is not necessary).

In Fig.~\ref{fig:5}(b), we plot the Fourier transform of the hole density as a function of the position along the molecular backbone. Strikingly, the figure shows that features corresponding to both $\omega_{\text{CM}}$ and $2\omega_{\text{CM}}$ are present across the entire length of the molecule, with varying strengths. Although the modulation at $2\omega_{CM}$ is largest near the central pair of carbon atoms, as discussed above, it is clearly also visible at both ends of the molecule. The similarity between the frequency-representation extracted from HHSS in Fig.~{\ref{fig:5}}(a) and the one extracted directly from the time-dependent hole density in (b) is very intriguing: we again observe two modulation contributions that are comparable in strength, with the $\omega_\text{CM}$ feature dominating slightly overall. Note also the clear variation of the two strengths with the harmonic energy in (a): around 45~eV the two components have comparable intensity while around 60~eV (35~eV) the fundamental (second harmonic) clearly dominates. This is a reflection of the multi-dimensionality of the CM dynamics itself, as also observed in the spatial variation in panel (b). In the harmonic signal, the CM dynamics may influence both the initial ionization and the final rescattering steps of the HHG process, depending on where the hole is localized along the molecular backbone in each of these steps. Because the ionization step alone would result in a global modulation of the $\omega_\text{CM}$ and/or $2\omega_\text{CM}$ components, we interpret the energy dependence of these components as evidence of a CM-induced modulation of the HHG signal associated with the rescattering step. Future investigation of this coherent CM modulation of the HHG process might also benefit from information contained in the harmonic phase, and/or some description of site-specific ionization and probing mechanisms.

\subsection{Experimental Viability}\label{sec:3C}

\begin{figure}[tb]
    \centering
    \includegraphics[width=\linewidth]{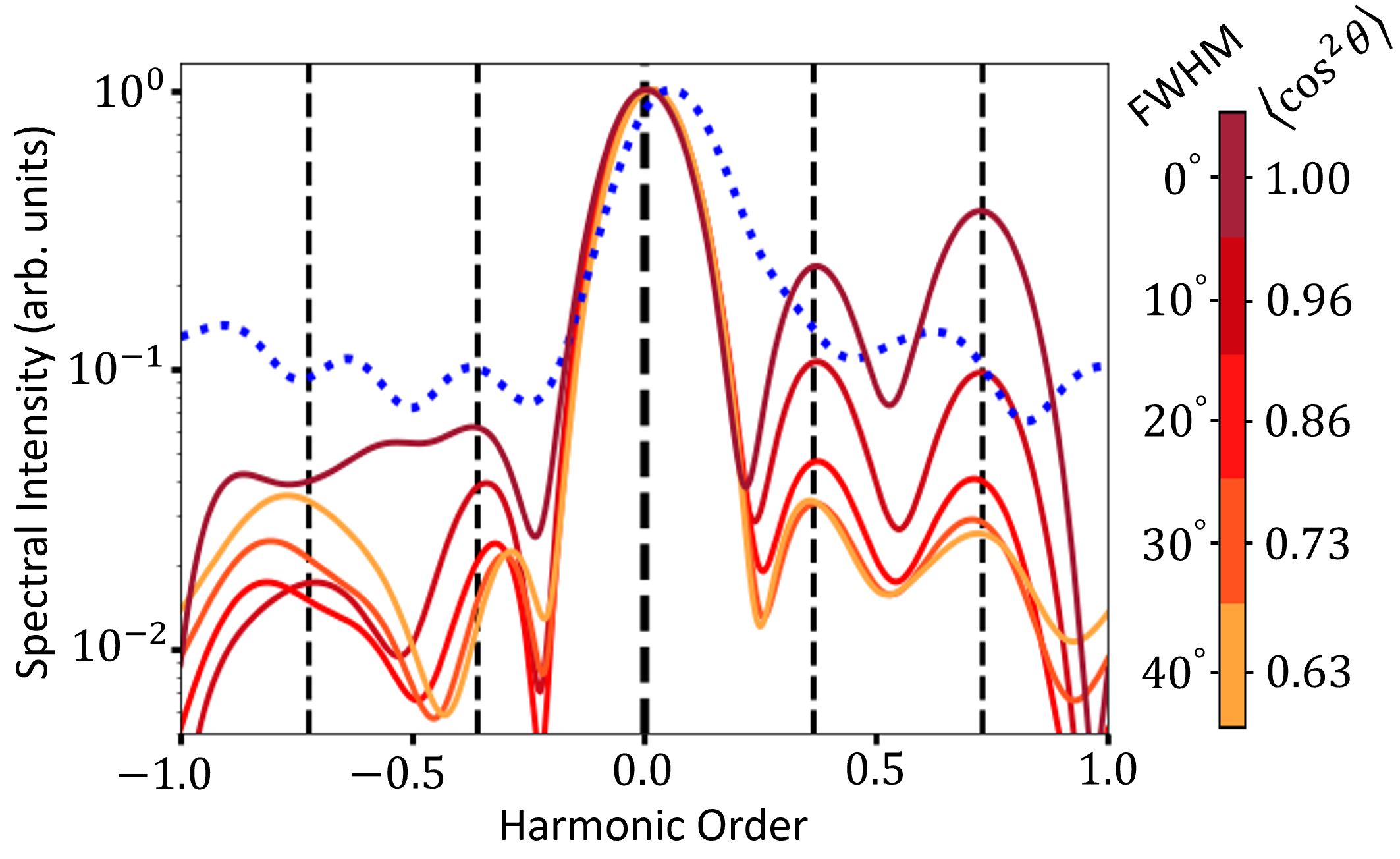}
    \caption{Effect of having a distribution of molecular-alignment angles around perpendicular alignment. The solid curves compare the stacked spectrum of Fig.~\ref{fig:2}~(a) when including various Gaussian angular distributions -- see colorbar. For comparison, the blue dotted curve shows the unaligned stacked spectrum. Vertical black lines mark the locations of the odd harmonic and sidebands.}
    \label{fig:6}
\end{figure}

Until now, we have assumed the idealized condition of a single, perfectly oriented, BrC$_4$H molecule undergoing CM. However, we have verified the viability of HHSS for more realistic experimental considerations. We start by investigating the effect of a distribution of molecular alignments. We use a Gaussian distribution of angles around perpendicular alignment and coherently average the corresponding CM+HHG dipole signals, for $\lambda=1575\, \text{nm}$. In Fig.~\ref{fig:6} we plot the evolution of the stacked spectrum when increasing the width of the alignment distribution from perfect orientation ($\theta \equiv 0^\circ$). Strikingly, up to 40$^\circ$ FWHM, we clearly see $\text{SB}_{1}$ and $\text{SB}_{2}$ sideband peaks, and we find that their positions can be located with error bars comparable to those in Fig.~\ref{fig:4}(b). The relatively lower strength of the sidebands when including the  distribution of alignment angles is caused by the the component of the MIR field that is parallel to the molecular axis which disturbs  the field-free CM dynamics. This generally leads to sidebands at slightly different frequencies for different alignment angles. For the completely unaligned sample (blue dotted curve in Fig.~\ref{fig:6}), the sidebands average out altogether.
We note that the CM induces different modulation amplitudes in the sub-cycle dipole signals of molecules with the same alignment but opposite orientations, except when the molecule is exactly perpendicular to the laser polarization (see also the discussion about finite-sampling effects in Appendix~\ref{sec:TDDFT_sims}). To generate Fig.~\ref{fig:6}, we therefore sample molecular-orientation angles covering a full 180$^\circ$ instead of the 90$^\circ$ range that would otherwise be sufficient for a linear molecule in a symmetric multi-cycle MIR field.

Next, we have found that the sideband signals are robust with respect to having only a fraction of the molecules in the HHG target gas undergoing CM. We consider a weighted average of the dipole signals from the target with and without CM -- thin red and solid blue curves in Fig.~\ref{fig:1}(c), respectively. We find that the intensity of the sidebands scales quadratically with the fraction of molecules with CM, as expected for a coherent process.  We also expect the sidebands to phase match to the same extent as the constituent harmonics in that pulse train, since they originate from the time-domain amplitude modulation of the individual bursts in the harmonic pulse train. 

Finally, we have tested the influence of a finite CM lifetime, {\it e.g.}, due to the onset of nuclear dynamics. When adding a phenomenological exponential decay to the CM-modulated dipole component, we still observe well-defined sidebands with a lifetime as short as 10~fs. Although the formal inclusion of nuclear motion is beyond the scope of this article, our preliminary calculations including (classical) nuclear dynamics described at the Ehrenfest level indicate that the linear, triple-bonded BrC$_4$H molecule is quite rigid, such that the localized hole motion is stable over multiple laser cycles.


\section{Summary and Outlook}

In summary, we have shown that periodic CM dynamics initiated in a BrC$_4$H molecule leads to a coherent time-dependent modulation of a time-delayed HHG probe, reflecting the sensitivity of the HHG process to the local charge density. This modulation is directly visible at the sub-laser-cycle level, as a function of the time delay, so that the Fourier transform of the delay-dependent harmonic spectrum contains information about the character of the CM motion. In the spectral domain and over several laser cycles, the migration dynamics manifest as sidebands whose energies correspond to the sum and the difference frequencies between the driving laser and the CM. These sidebands constitute a background-free probe of the CM motion from which one can unambiguously extract the migration frequencies by scanning the laser wavelength. Interestingly, a localized, particle-like CM motion in BrC$_{4}$H is characterized by multiple sidebands, associated with $\omega_{\text{CM}}$ and $2\omega_{\text{CM}}$. We note that the same type of periodic, particle-like CM modes are commonly found in other conjugated halocarbons~\cite{folorunso2021}, given a proper initial localized-ionization condition. This generality opens the door to adapting HHSS to other classes of molecules. Looking forward, we note that HHG has a built-in ``clock,'' with sub-cycle resolution, that can be accessed through the measurement of the phases of the harmonics (attochirp)~\cite{paul2001, mairesse2003}. Therefore, a full characterization of the HHSS harmonics/sidebands offers a promising path to view the CM motion with sub-cycle resolution.


{\bf Data availability}: The TDDFT simulation data and Python scripts we use to produce the figures are available at [authors will provide link to public repository for production].


\begin{acknowledgements}

This work was supported by the U.S.\ Department of Energy, Office of Science, Office of Basic Energy Sciences, under Award No.~DE-SC0012462.
Portions of this research were conducted with high performance computational resources provided by Louisiana State University (\url{http://www.hpc.lsu.edu}) and the Louisiana Optical Network Infrastructure (\url{http://www.loni.org}).

\end{acknowledgements}


\appendix

\section{TDDFT simulations} \label{sec:TDDFT_sims}

In order to describe both the high-harmonic generation (HHG) and charge-migration (CM) processes, we use grid-based TDDFT \cite{kohn1965} with a local density approximation exchange-correlation functional \cite{Bloch1929, dirac1930, perdew1981, marques2012} and an average-density self-interaction correction \cite{legrand2002} within Octopus \cite{andrade2012, andrade2015}.
Following the blueprint of~\cite{hamer2021}, we put the BrC$_4$H molecule at the center of the simulation box with the $x$ axis parallel to the laser-polarization direction. We then vary molecular orientation by rotating the target inside that box.
For each TDDFT simulation, we scale the length of the simulation box in the $x$ (field) direction with the quiver radius (twice the quiver radius plus 10~a.u.\ buffer end-to-end), to improve the selection of the HHG-short-trajectory contribution (see below).
In the transverse direction that contains the molecular backbone in the ideal case of perfect alignment ($z$-axis) we set the box length to 90 a.u., and to 40 a.u.\ in the third direction ($y$-axis). 
In all directions we use a discretization-grid spacing of 0.3~a.u., and we include a complex absorbing potential (CAP) that extends 15 a.u.\ from the edges of the box. 
To propagate the TDDFT dynamics, we use the enforced time-reversal symmetry (ETRS) scheme in Octopus~\cite{castro2004} with a time step of 0.05 atomic units.

For harmonic computations, we use a combination of an MIR field, with a peak intensity of $6 \times 10^{13}\ \text{W}/\text{cm}^{2}$, and a weak attosecond pulse train (APT)~\cite{schafer2004} consisting of the 9$^\text{th}$-to-17$^\text{th}$ odd harmonics of te MIR laser, and having 2\% of the MIR field amplitude. The timing of the APT, 0.08 optical cycles after the extrema of the MIR electric field, is set to enhance short trajectories~\cite{schafer2004,hamer2021}. 
Finally, we ramp up the MIR+APT field with a $\sin^{2}$ envelope for two full optical cycles (followed by a constant-amplitude envelope), and control its delay with respect to the initiation of the CM dynamics.
Note that the sub-cycle control of the MIR-CM delay is necessary only for delay scans like in Fig.~\ref{fig:5}. For multi-cycle HHG spectra, because the CM and laser frequencies are incommensurate, only the amplitude (not the frequency) of the sidebands is affected by that delay.

\begin{figure}[tb]
    \centering
    \includegraphics[width=\linewidth]{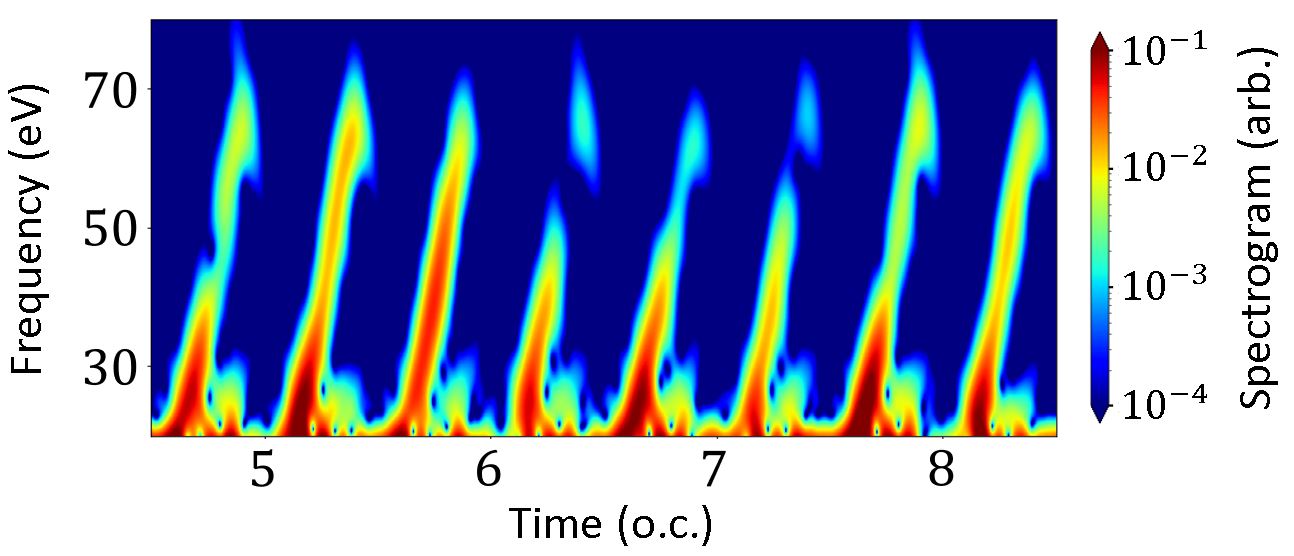}
    \caption{Gabor transform for the HHSS signal of Fig.~1~(c) in the main text for 1575~nm.}
    \label{fig:A1}
\end{figure}

In Fig.~\ref{fig:A1}, we show the Gabor transform of the dipole-acceleration signal associated with the HHSS of Fig.~1~(c), for an MIR wavelength of 1575~nm. Every half laser cycle, the Gabor transform shows a clear enhancement of the short-trajectory signal and (comparatively) no long trajectories. The differences in the successive short-trajectory branches reflect the modulation of the HHG signal induced by the CM dynamics.
In turn, the finite number of attosecond bursts that contribute to each harmonic and sideband (vertical slices in the figure) leads to finite-sampling effects that affect their relative strengths in HHG spectra.

\section{Localized-hole initial condition} \label{sec:init_cond}

To initiate the particle-like CM dynamics in BrC$_4$H, we create a one-electron valence hole that is localized on the Br end of the molecule. The resulting excited molecular cation is far from the DFT-ground-state equilibrium and not readily accessible from Octopus' self-consistent-field algorithms. Instead, we reproduce the initial localized hole of Ref.~\cite{folorunso2021}, obtained with constrained DFT (cDFT)~\cite{eshuis2009} in NWChem~\cite{lopata2011, apra2020}, by taking a linear combination of the ground-state molecular orbitals of BrC$_4$H$^+$ (from Octopus).

Specifically, we use the \texttt{TransformStates} block in Octopus, which transforms the initial ground-state wavefunctions at $t=0$ using a given transformation matrix.
We determine the coefficients for the transformation matrix by calculating the spatial overlap between cDFT orbitals from NWChem, {\it i.e.}, with a Br-localized hole, and ground-state DFT orbitals of BrC$_4$H$^+$ from Octopus. We impose the transformation matrix to be unitary such that, at the beginning of our TDDFT simulations of CM, the Kohn-Sham orbitals remain a set of orthonormal one-electron wave functions. With this method, we reproduce the cDFT-localized hole from NWChem with more than 99\% accuracy ($>$99\% overlap between NWChem's c-DFT Kohn-Sham orbitals and the reconstructed ones in Octopus).
More importantly, with this initial hole, we are able to trigger periodic CM dynamics, with a clear particle-like hole that travels back and forth through the molecule \cite{folorunso2021} -- see Fig.~1(b)  -- which we study with HHSS.
This particle-like hole motion is consistent with previous studies on the effect of the precise initial condition on CM, which have found that the CM dynamics are nearly unchanged for a large range of parameters~\cite{Mauger2022}.

\section{HHG and HHSS computations} \label{sec:HHG_spectra}

We compute all HHG/HHSS spectra from the Fourier transform of the total dipole acceleration, including a $\cos^2$ window function to avoid numerical artifacts in the Fourier transform (because of the lack of periodicity of the dipole signal).
When computing HHSS spectra, like in Fig.~1(c), we select a part of the dipole signal where the driving-field amplitude is constant.
We discard the ramp-up part of the pulse where the smaller instantaneous field amplitude means that trajectories have an effectively smaller quiver radius and thus long-trajectory contributions are not removed by the CAP (see section~\ref{sec:TDDFT_sims}). This leads to a residual long-trajectory component in our HHSS signal that is normally not seen in experiments.
We have checked that including the ramp up has only cosmetic effects on HHSS spectra and does not change the conclusion of our sideband analyses.


\end{document}